# Edge Dislocations Can Control Yield Strength in Refractory Body-Centered-Cubic High Entropy Alloys

Francesco Maresca[1,2*,§], Chanho Lee[3,§], Rui Feng[3], Yi Chou[4], T. Ungar[5], Michael Widom[6], Ke An[7], John D. Poplawsky[8], Yi-Chia Chou[4], Peter K. Liaw[3], and W. A. Curtin[1]

1. Laboratory for Multiscale Mechanics Modeling, École Polytechnique Fédérale de Lausanne, CH-1015 Lausanne, Switzerland
2. Faculty of Science and Engineering, University of Groningen, Groningen, 9474AG, Netherlands
3. Department of Materials Science and Engineering, The University of Tennessee, Knoxville, TN 37996-2100, USA
4. Department of Electrophysics, National Chiao Tung University, Hsinchu, 30010, Taiwan
5. Department of Materials Physics, Eötvös University, Budapest, P.O. Box 32, H-1518, Hungary
6. Department of Physics, Carnegie Mellon University, Pittsburgh, PA 15213, USA
7. Neutron Scattering Division, Oak Ridge National Laboratory, Oak Ridge, TN, 37831, USA
8. Center for Nano-phase Materials Sciences, Oak Ridge National Laboratory, Oak Ridge, TN, 37831, USA

*Corresponding author: f.maresca@rug.nl

§These authors contributed equally to this work

**Energy efficiency is motivating the search for new high-temperature metals. Some new body-centered-cubic random multicomponent ``high entropy alloys (HEAs)'' based on refractory elements (Cr-Mo-Nb-Ta-V-W-Hf-Ti-Zr) possess exceptional strengths at high temperatures but the physical origins of this outstanding behavior are not known. Here we show, using integrated neutron-diffraction (ND), high-resolution transmission electron microscopy (HRTEM), and theory, that the high strength and strength retention of a NbTaVTi alloy and a new high-strength/low-density CrMoNbV alloy are attributable to edge dislocations. This is surprising because plastic-flow in BCC elemental metals and dilute alloys is universally accepted to be controlled by screw dislocations. We use the insight and theory to perform a computationally-guided search over $10^7$ BCC HEAs and identify over $10^6$ possible ultra-strong high-temperature alloy compositions for future exploration.**

Achieving the urgent societal goals of reduced emissions and increasing energy efficiency is driving the development of new materials. One path is lightweight materials (Mg, Al, reinforced plastics) *(1,2,3)* for low-temperature applications, such as transportation, while a second path is high-temperature damage-tolerant materials for increased combustion efficiency (superalloys, TiAl) *(4,5)*. The new ``high entropy alloys (HEAs)'' are single-phase crystalline materials with many components that randomly occupy the atomic sites of the crystal lattice *(6-11)*. HEAs can have remarkable yield strength, ductility, and/or fracture toughness. The two body-centered-cubic (BCC) HEAs MoNbTaVW and MoNbTaW retain high strength at temperatures far above those for existing superalloys (Figure 1) *(9,11,12)* but the mechanisms enabling this performance are not well-established *(11,12)*. Here, we show that the high strength and strength retention of both the recent NbTaVTi and new CrMoNbV HEAs (Figure 1) are controlled by edge dislocations. Our findings are unexpected because screw dislocations are widely understood to control plastic flow in BCC elemental metals and dilute alloys *(13,14)*. However, unlike in dilute alloys, our recent theory shows that edge dislocations in some complex HEAs can encounter very large energy barriers to glide *(15)*, and hence high strength and strength retention at high temperatures. Thus, while the new CrMoNbV alloy has the highest retained strength to date at T = 1,173K, a theory-guided search over the entire Cr-Mo-Nb-V-W-Ti-Hf-Zr-Al composition space predicts over 1,000,000 new alloys with even better performance.

Alloys of nominal compositions of NbTaTiV and CrMoNbV were synthesized, and in-situ neutron-diffraction (ND) measurements were performed (Materials and Methods; S.M. 1, 2; Supp. Fig. 1). The ND patterns show a single BCC solid solution at all temperatures up to 1,173 K (Figures 2a, 3a). Atom probe tomography reveals NbTaTiV to contain 1.15 at% (atomic percent) O and 0.45 at% N while CrMoNbV has low interstitial contents (0.083 at% O, 0.034 at% N, and 0.062 at%C). CrMoNbV also has an attractive density ($\rho$ = 8.08 g/cm$^3$) and melting point ($T_m \sim$ 2,134 K), and is studied in the as-cast state. Figure 1 shows the measured yield strengths $\sigma_y$ versus temperature at a strain rate of $1 \times 10^{-3}$ s$^{-1}$; high strength is retained up to 1,173 K (Supp. Figs. 2,3). The strength of NbTaTiV is comparable to MoNbTaW and MoNbTaWV, which have been novel among BCC HEAs for their exceptional high-T strength retention *(11,12)* and have been predicted to be controlled by edge dislocations *(15)*. Moreover, the CrMoNbV alloy – never tested before – has the highest reported strength at homologous temperature $T/T_m$ = 0.55 *(12)*.

We now demonstrate that edge dislocations, not screw dislocations, control the plasticity in both NbTaTiV and CrMoNbV. First, TEM analysis on NbTaTiV shows that the dislocations have Burgers vectors of the <111>a/2 type typical of BCC elements and dilute alloys (Supp. Fig. 4; S.M. 3); further TEM analysis also suggests that edge dislocations predominate. Second, line broadening of the ND peaks during deformation is due to inhomogeneous strain fields generated by mechanically-induced substructures, including dislocations. For a given {hkl} lattice plane in the crystal, ND yields the evolution of both the interplanar spacing, $d_{hkl}$, in the elastic regime and the peak broadening (full width at half maximum, $\Delta d_{hkl}$) in the plastic regime. Figures 2b,3b present the measured axial and transverse lattice strains versus applied stress at T = 293 K. The planar Young's modulus, $E_{hkl}$, and Poisson ratio, $v_{hkl}$, are derived using Kroner's model, and the cubic elastic constants, $C_{11}$, $C_{12}$, and $C_{44}$ are then computed (Supp. Fig. 5, S.M. 4). The resulting Zener anisotropies, $\frac{2C_{44}}{C_{11}-C_{12}} = 1.2405$ (NbTaTiV) and 0.71 (CrMoNbV), a regime where peak broadening is particularly sensitive to dislocation character (see below). The observed peak broadening during subsequent deformation is analyzed using the quantitative state-of-the-art CMWP method *(16)*. In addition to instrumental response, HEAs have a peak broadening in the undeformed state due to local deviations of atoms from the perfect lattice positions; it is important that this initial broadening be subtracted from the mean-square broadening at finite deformation to isolate the role of the dislocations. The line broadening by dislocations is related to the dislocation contrast factor for each {hkl} plane via the q parameter as

$$C_{hkl} = C_{h00}\left[1 - q\left(\frac{h^2k^2+k^2l^2+h^2l^2}{h^2+k^2+l^2}\right)\right] \qquad (2)$$

where *q* is a function only of the dislocation Burgers vector, character, and elastic anisotropy. The CWMP analysis performed on NbTaTiV samples, for <111>a/2 dislocations and deformed to plastic strains of 1.8, 6.8, and 11.8%, reveals *q* values as shown in Figure 2c. These are all consistent with the theoretical value of -0.6 < *q* < -0.4 (accounting for uncertainty in the elastic constants) for edge dislocations, and far differ in both magnitude and sign than the range +1.9 < *q* < +2.1 for screw dislocations. The simpler modified Williamson-Hall (mWH) analysis *(17)* of $\Delta K_{hkl} = -\frac{\Delta d_{hkl}}{d_{hkl}^2}$ versus $K = \frac{1}{d}$ as $(\Delta K_{hkl})^2 = (0.9/D)^2 + BK_{hkl}^2 C_{hkl}$ ($B$ a constant) yields the same conclusion (Figure 2d). Similarly, CMWP and mWH analyses for the CrMoNbV alloy at plastic strains of 0.8%, 4.2% and 5.3% yield $q$ values shown in Figure 3c,d which can be compared to the theoretical edge value of -2 and the theoretical screw value of +1.5. The quality of the neutron pattern for CrMoNbV is lower than for NbTaTiV, especially at the very low plastic strain of 0.8%, making that result less reliable. Overall, the neutron data demonstrates that edge dislocations are more dominant than screw in both NbTaTiV and CrMoNbV.

The significant role of edge dislocations in NbTaTiV and CrMoNbV is further supported by Annular Bright-Field (ABF)-STEM and Stereographic analysis of specimens deformed at T = 293 K (Materials and Methods). Figures 2e,3e show the ABF-STEM images of the dislocation networks in NbTaTiV and CrMoNbV, at plastic strains of 11.8% and 4.2%, respectively, taken slightly off the [110] zone axis to enhance the contrast. All dislocations having a line of length over 5 nm are identified (blue and red lines). The identified line direction is compared with the stereographic projection, and if the difference is less than 5 degrees then the dislocation character is determined. If two characters are possible, the one with the closest match to the dislocation line is chosen. If the dislocation line does not meet these criteria, it is not considered (e.g., the finer-scale dislocation tangles in the images). Figures 2f,3f show the results of the stereographic projection analysis, which reveals the characters of the long straight dislocation lines; the observed dislocations (blue and red in Figures 2e,3e) are highlighted in bold and lie along the lines shown. In NbTaTiV, the measured dislocations are predominantly of edge character (58 of 77, or 75%). In CrMoNbV, the dislocations are again predominantly edge (23 of 41, or 56%). Dislocation contrast, i.e. **g·b** analysis, can also reveal dislocation character for some cases. Analysis shown in S.M. 3 further demonstrates the dominance of edge dislocations in NbTaTiV.

The observations of a high fraction of edge dislocations relative to screw dislocations is in distinct contrast to typical studies in BCC metals and other HEAs *(18)* that almost exclusively show long straight screw dislocations and do not have good high temperature strength. While results from TEM and neutron diffraction are not quantitatively identical, both results consistently show that edge dislocations are prevalent and dominant, as compared to screws.

Our experimental demonstrations that plasticity in a BCC alloy can be controlled by edge dislocations is very surprising because it is counter to the nearly-universally-accepted understanding, based on many observations in elemental and dilute BCC alloys over many decades, that screw dislocations completely dominate the plastic flow behavior *(13,14)*.

To further cement that the edge dislocations control the strength, we apply a new theory for the yield strength versus temperature and strain rate for edge dislocations moving through a random BCC alloy *(15)*. In the theory, each alloying element is viewed as a solute that interacts with a dislocation in a hypothetical homogeneous "average" alloy that has all of the macroscopic properties of the true random alloy. Fluctuations in the local arrangements of solutes create large local variations in the potential energy for the dislocation. The dislocation thus spontaneously adopts a low-energy wavy structure to take advantage of the low-energy solute environments and avoid the high-energy environments. Plastic flow then requires the temperature- and stress-assisted thermal activation of the dislocations out of the low-energy environments and over the large barriers created by the adjacent high-energy environments along the glide plane in the random alloy. The full theory is reduced to an analytic model by using the elasticity approximation $U^i(x,y) = -p(x,y)\Delta V_i$ for the solute/dislocation interaction $U^i(x,y)$ of a solute of the type *i* at position *(x, y)* under the pressure field *p(x,y)* due to a dislocation lying along *z* and centered at the origin, where $\Delta V_i$ is the misfit volume of the type $i$ solute in the average alloy. The yield stress as a function of temperature and strain rate $\dot{\varepsilon}$ is then *(15)*

$$\sigma_y(T, \dot{\varepsilon}) = \sigma_{y0} \left[1 - \left(\frac{kT}{\Delta E_{b0}} \ln \frac{\dot{\varepsilon}_0}{\dot{\varepsilon}}\right)^{2/3}\right] , \frac{\sigma_y}{\sigma_{y0}} \geq 0.5 \qquad [3a]$$

$$\sigma_y(T, \dot{\varepsilon}) = \sigma_{y0} \exp\left(-\frac{1}{0.55} \frac{kT}{\Delta E_{b0}} \ln \frac{\dot{\varepsilon}_0}{\dot{\varepsilon}}\right) , \frac{\sigma_y}{\sigma_{y0}} < 0.5 \qquad [3b]$$

where the zero-temperature yield stress $\sigma_{y0}$ and zero-stress energy barrier $\Delta E_{b0}$ are computed as

$$\sigma_{y0} = 3.067 \, A_\sigma \, \alpha^{-1/3} \mu \left(\frac{1+\nu}{1-\nu}\right)^{4/3} \left(\sum_i \frac{c_i \Delta V_i^2}{b^6}\right)^{2/3} \qquad [4a]$$

$$\Delta E_{b0} = A_E \, \alpha^{1/3} \mu b^3 \left(\frac{1+\nu}{1-\nu}\right)^{2/3} \left(\sum_i \frac{c_i \Delta V_i^2}{b^6}\right)^{1/3} \qquad [4b]$$

Here, $\mu$ and $\nu$ are the alloy elastic constants, $\{c_i\}$ the solute concentrations, $\alpha = 1/12$ a line tension parameter, and $\dot{\varepsilon}_0 = 10^4 s^{-1}$ a reference strain rate. The coefficients $A_\sigma$ and $A_E$ are computed in the full theory for each alloy composition. *There are thus no fitting parameters in the theory*, only material properties (elastic constants, misfit volumes, and line tension) and alloy composition. However, the coefficients fall in a narrow range of $A_\sigma = 0.040 \pm 0.004$ and $A_E = 2.00 \pm 0.2$ across a wide spectrum of BCC HEAs, indicating that the dominant material properties are the misfit volumes and elastic constants.

Applying the theory to NbTaTiV, we first compute the misfit volumes of all solutes using Vegard's Law, which accurately predicts alloy atomic volumes of refractory BCC HEAs studied to date *(9)*. The BCC atomic volumes $\{V_{0i}\}$ are Nb=17.952, Ta=17.985, Ti=17.387, and V=14.02 in Å$^3$, respectively, the alloy volume is $V = \sum_i c_i V_{0i}$, and the misfit volume for solute $i$ is $\Delta V_i = V_{0i} - V$. The alloy elastic constants are computed using the rule of mixtures for the elemental $C_{11}$, $C_{12}$ and $C_{44}$ to obtain the bulk modulus $B = \frac{C_{11}+2C_{12}}{3}$ and $\mu = \sqrt{C_{44}(C_{11}-C_{12})/2} = 47.8 \, GPa$, and then $\nu = \frac{3B-2\mu}{2(3B+\mu)} = 0.365$, consistent with experiments (see Supp. Fig. 5). We use the precise computed coefficients $A_\sigma = 0.0437$ and $A_E = 2.02$ from the full theory for the NbTaTiV alloy. The predicted strength versus temperature for NbTaTiV *with no interstitial content* at the experimental strain rate of $10^{-3}$/s is shown in Figure 4a along with the present experiments. The theory trend is good but much lower than the present experiments, although comparable to literature data at T = 300 K. We attribute this difference to the 1.6 at% O + N impurities in our alloy because it is known that 2 at% O or N strengthens a BCC HEA by ~ 400 - 500 MPa at T = 293 K (Figure 4a) *(19)*. Furthermore, the tetragonal misfit strains for O and N in Nb, Ta, and V, as computed via DFT, are all large ( $\varepsilon_{11}$ ~ 0.60; $\varepsilon_{22} = \varepsilon_{33}$ ~ − 0.1 ), consistent with high strengthening. Semi-quantitatively, using a new Nb-O interatomic potential *(20)* to compute the interaction of O with an *edge* dislocation in Nb and assuming the same interaction energies for the NbTaVTi alloy, the theory can be extended to include the addition of 1.6 at% O interstitials and predicts an increase in strength of ~ 300 MPa at T = 300 K, reaching reasonable agreement with experiment (Figure 4a).

We next apply the theory to the new CrMoNbV alloy. Comparison with experiments is facilitated by the very low interstitial content, as compared to NbTaTiV. Equations 4a,b show that the solute misfit volumes are the critical component for strengthening, and Cr has a small BCC atomic volume of 12.321 Å$^3$ and so should lead to high strengths. We apply Eqs. 3-4 using (i) Vegard's law, (ii) the additional atomic volumes of Cr=12.321 Å$^3$ and Mo=15.524 Å$^3$, (iii) alloy elastic moduli of $\mu = 86.24$ GPa and $\nu = 0.3$, and (iv) the precise computed coefficients $A_\sigma = 0.0344$ and $A_E = 1.93$ for CrMoNbV. The predicted yield strength versus temperature for CrMoNbV is shown in Figure 4a and agrees well with the measured strengths. The predicted and measured strengths of this new alloy exceed those of *all* previously-reported single-phase BCC HEA alloys at temperature of 1,173 K *(11,12,21)* and at the homologous temperature T/T$_m$ = 0.55 (see *(12)*). Our predictions based on the edge dislocation are fully consistent with their significant presence as revealed by both neutron diffraction and ABF-STEM.

With the new understanding of the key role of edge dislocations in the strengthening of BCC HEAs, especially at high temperatures, we can now identify new promising alloy compositions. We note that screw dislocations remain important in some BCC HEAs, but the theoretical mechanisms fail at higher T and accurate theory inputs are difficult to obtain *(22-23)*. Thus, a computationally-guided search for new high-performance alloys based on edge-dislocation strengthening is a *computable* mechanistic path for the design and discovery of further new alloys with high strengths, high-temperature strength retention, and low density. To this end, we use Eqs. 3-4, the average coefficients of $A_\sigma = 0.040$ and $A_E = 2.00$, Vegard's Law, and the rule-of-mixtures for elastic constants, to search across more than 10,000,000 compositions in the 10-component Cr-Mo-Nb-Ta-V-W-Hf-Ti-Zr-Al family (S.M. 5). As shown in Figure 4b,c, we find ~ 6,000,000 alloys with estimated strengths over 1 GPa *at T = 1,300 K*, and ~ 1,300,000 over 2 GPa, far exceeding the strengths of any existing alloys. *At 1,300 K*, many alloys also have strength/density > 0.25 GPa g/cm$^3$ that is the highest achieved to date *at room temperature*. We propose two super-high-strength/high-T alloys of $Mo_5W_{2.5}CrZrHf$ and $Mo_{2.5}W_{2.5}CrZrHf$ for fabrication and testing, with predicted strengths of ~ 3 GPa at 1,300 K and considering preliminary thermodynamic assessment (S.M. 6) that suggest limiting Cr-Zr-Hf content to avoid intermetallics. We also propose two super-high-strength/high-T alloys $Mo_6WCrZrHf$ and $Mo_{2.5}TaWV_{2.5}CrZrHf$ with limited W yet predicted strengths of ~ 2.5 GPa at 1,300 K. Future combinations of our design strategy with detailed thermodynamics *(24-25)* and added constraints (e.g. high ductility criterion) may lead to the discovery of new alloys that can achieve the multi-objective performance required in many critical engineering applications.

**References**

1. Z. Wu, R. Ahmad, B. Yin, S. Sandlöbes, W.A. Curtin, Mechanistic origin and prediction of enhanced plasticity in magnesium alloys. *Science* **26**, 447-452 (2018).

2. J.H. Martin, B.D. Yahata, J.M. Hundley, J.A. Mayer, T.A. Schaedler, T.M. Pollock, 3D printing of high-strength aluminium alloys. *Nature* **549**, 365-369 (2017).

3. M.F.L. Volder, S.H. Tawfick, R.H. Baughman, A.J. Hart, Carbon nanotubes: present and future commercial applications. *Science* **339**, 535-539 (2013).

4. T.M. Smith, B.D. Esser, N. Antolin, A. Carlsson, R.E.A. Williams, A. Wessman, T. Hanlon, H.L. Fraser, W. Windl, D.W. McComb, M.J. Mills, Phase transformation strengthening of high-temperature superalloys. *Nature Commun.* **7**, 13434 (2016).

5. G. Chen, Y. Peng, G. Zheng, Z. Qi, M. Wang, H. Yu, C. Dong, C.T. Liu, Polysynthetic twinned TiAl single crystals for high-temperature applications. *Nature Mater.* **15**, 876-881 (2016).

6. B. Gludovatz, A. Hohenwarter, D. Catoor, E.H. Chang, E.P. George, R.O. Ritchie, A fracture-resistant high-entropy alloy for cryogenic applications. *Science* **345**, 1153-1158 (2014).

7. Z. Wu, H. Bei, G. Pharr, E. George, Temperature dependence of the mechanical properties of equiatomic solid solution alloys with face-centered cubic crystal structures. *Acta Mater.* **81**, 428-441 (2014).

8. O.N. Senkov, G.B. Wilks, J.M. Scott, D.B. Miracle, Mechanical properties of Nb25Mo25Ta25W25 and V20Nb20Mo20Ta20W20 refractory high entropy alloys. *Intermetallics* **19**, 698-706 (2011).

9. H.W. Yao, J.W. Qiao, M.C. Gao, J.A. Hawk, S.G. Ma, H.F. Zhou, Y. Zhang, NbTaV-(Ti,W) refractory high-entropy alloys: Experiments and modeling. *Mater. Sci. Eng. A* **674**, 203-211 (2016).

10. Z. Li, K.G. Pradeep, Y. Deng, D. Raabe, C.C. Tasan, Metastable high-entropy dual-phase alloys overcome the strength–ductility trade-off. *Nature* **534**, 227-230 (2016).

11. O.N. Senkov, D.B. Miracle, K.J. Chaput, J.-Ph. Couzinie, Development and exploration of refractory high entropy alloys-A review. *J. Mater. Res.* **33**, 3092-3128 (2018).

12. O.N. Senkov, S. Gorsse, D.B. Miracle, High temperature strength of refractory complex concentrated alloys, *Acta Mater.* **175**, 394-504 (2019).

13. D. Rodney, J. Bonneville, Dislocations. *In Physical Metallurgy*, Elsevier Oxford (2014).

14. D.R. Trinkle, C. Woodward, The chemistry of deformation: How solutes soften pure metals. *Science* **310**, 1665-1667 (2005).

15. F. Maresca, W.A. Curtin, Mechanistic origin of high strength in refractory BCC high entropy alloys up to 1900K, *Acta Mater.* **182**, 235-249 (2020).

16. G. Ribárik, T. Ungár, J. Gubicza, MWP-fit: a program for multiple whole-profile fitting of diffraction peak profiles by ab initio theoretical functions. *J. Appl. Crystallogr.* **34**, 669-676 (2001).

17. T. Ungár, I. Dragomir, Á. Révész, A. Borbély, The contrast factors of dislocations in cubic crystals: the dislocation model of strain anisotropy in practice. *J. Appl. Crystallogr.* **32**, 992-1002 (1999).

18. J.-Ph. Couzinié, L. Lilensten, Y. Champion, G. Dirras, L. Perrière, I. Guillot, On the room temperature deformation mechanisms of a TiZrHfNbTa refractory high-entropy alloy. *Mater. Sci. Eng. A* **645**, 255-263 (2015).

19. Z. Lei, X. Liu, Y. Wu, H. Wang, S. Jiang, S. Wang, X. Hui, Y. Wu, B. Gault, P. Kontis, D. Raabe, L. Gu, Q. Zhang, H. Chen, H. Wang, J. Liu, K. An, Q. Zeng, T.-G. Nieh, Z. Lu, Enhanced strength and ductility in a high-entropy alloy via ordered oxygen complexes. *Nature* **563**, 546-550 (2018).

20. P.-J. Yang, Q.-J. Li, T. Tsuru, S. Ogata, J.-W. Zhang, H.-W. Sheng, Z.-W. Shan, G. Sha, W.-Z. Han, J. Li, E. Ma, Mechanism of hardening and damage initiation in oxygen embrittlement of body-centered-cubic niobium. *Acta Mater.* **168**, 331-342 (2019).

21. F.G. Coury, M. Kaufman, A.J. Clarke, Solid-solution strengthening in refractory high entropy alloys. *Acta Mater.* **175**, 66-81 (2019).

22. S. I. Rao, E. Antillon, C. Woodward, B. Akdim, T.A. Parthasarathy, O.N. Senkov, Solution hardening in body-centered cubic quaternary alloys interpreted using Suzuki's kink-solute interaction model. *Scripta Mater.* **165**, 103-106 (2019).

23. F. Maresca, W.A. Curtin, Theory of screw dislocation strengthening in random BCC alloys from dilute to "High-Entropy" alloys. *Acta Mater.* **182**, 144-162 (2020).

24. R. Feng, P.K. Liaw, M.C. Gao, M. Widom, First-principles prediction of high-entropy-alloy stability. *npj Comput. Mater.* **3**, 50 (2017).

25. Y. Ikeda, B. Grabowski, F. Körmann, Ab initio phase stabilities and mechanical properties of multicomponent alloys: A comprehensive review for high entropy alloys and compositionally complex alloys. *Mater. Charact.* **147**, 464–511 (2019).

**Acknowledgments**

The authors thank Michael C. Gao for help in the design of the single-phase refractory HEAs using CALculation of PHAse Diagram (CALPHAD) and Gian Song and Hahn Choo for help with the analysis of the in-situ neutron diffraction data. The authors thank Dr. Binglun Yin for providing the DFT computation of the O and N misfit strains in Nb, Ta and V.

**Funding:** CL, RF, and PKL thank the U.S. Army Research Office for support of the present work through projects, W911NF-13-1-0438 and W911NF-19-2-0049. PKL thanks the National Science Foundation for the support of the present work through projects, DMR-1611180 and 1809640. CL acknowledges the partial support from the Center of Materials Processing, a Tennessee Higher Education Commission (THEC) Center of Excellence located at The University of Tennessee,

Knoxville. FM and WAC acknowledge the partial support for the current work from the European Research Commission Advanced Grant, "Predictive Computational Metallurgy", ERC Grant agreement No. 339081 - PreCoMet. FM and WAC also thank Prof. H. Sheng and Prof. E. Ma for sharing their Nb-O interatomic potential. YC and YCC thank the funding from the Ministry of Science and Technology (MOST) of Taiwan under Grant No. MOST-107-2636-M-009-002, the core facility support at National Chiao Tung University (NCTU) from MOST. YCC thanks the partial support from the Ministry of Education (MOE) through the SPROUT Project - Center for Smart Semiconductor Technologies of NCTU, Taiwan, and Center for the Semiconductor Technology Research from The Featured Areas Research Center Program within the framework of the Higher Education Sprout Project by the MOE in Taiwan, and the MOST, in Taiwan, under Grant MOST-108-3017-F-009-003. Research at the Spallation Neutron Source at the Oak Ridge National Laboratory (ORNL) was partially sponsored by the Scientific User Facilities Division, Office of Basic Energy Sciences, U.S. Department of Energy. The APT experiments were performed as a part of a user proposal at the Center for the Nanophase Materials Science (CNMS), ORNL, which is a DOE Office of Science User Facility.

**Competing interests**: The authors declare that they have no competing financial interests.

**Data and materials availability:** For access to more detailed data than are given in the article or the Supplementary Materials, please contact the authors.

**Supplementary Materials**

Materials and Methods

Supplementary Text

Table S1-S2

Fig S1-S5

References (26-39)

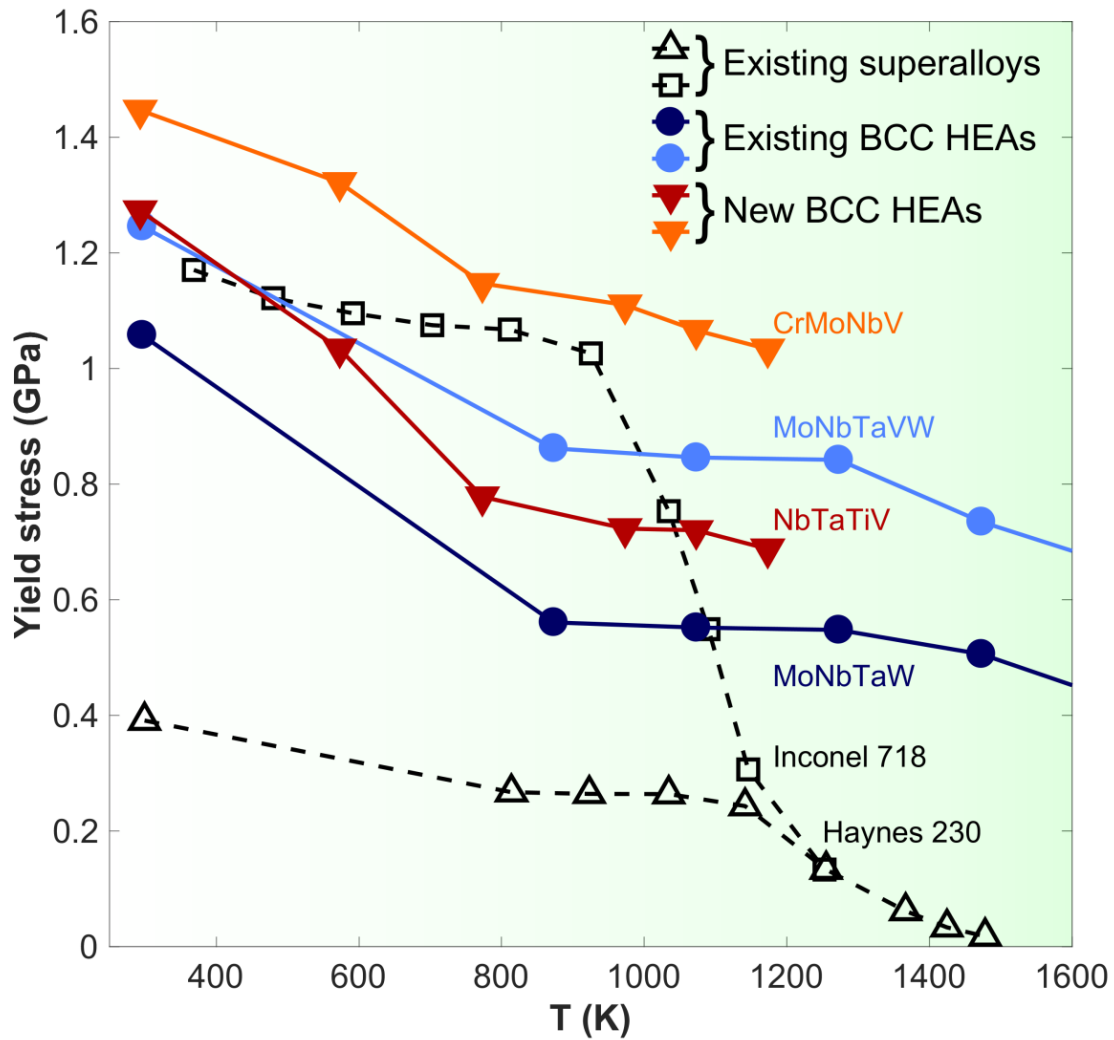

**Figure 1. Alloy strengths *vs* temperature.** The refractory BCC HEAs retain high strengths up to temperatures well beyond those where superalloys lose strength. The NbTaTiV alloy here is comparable to the literature MoNbTaVW and MoNbTaW alloys. The strength of the new CrMoNbV alloy, predicted to be even stronger, significantly exceeds the strengths of the existing BCC HEAs. The strengths of NbTaTiV and CrMoNbV are found here to be controlled by edge dislocations – see Figures 2, 3, and 4 – not screw dislocations. Literature data are reproduced from Ref. *(8)*.

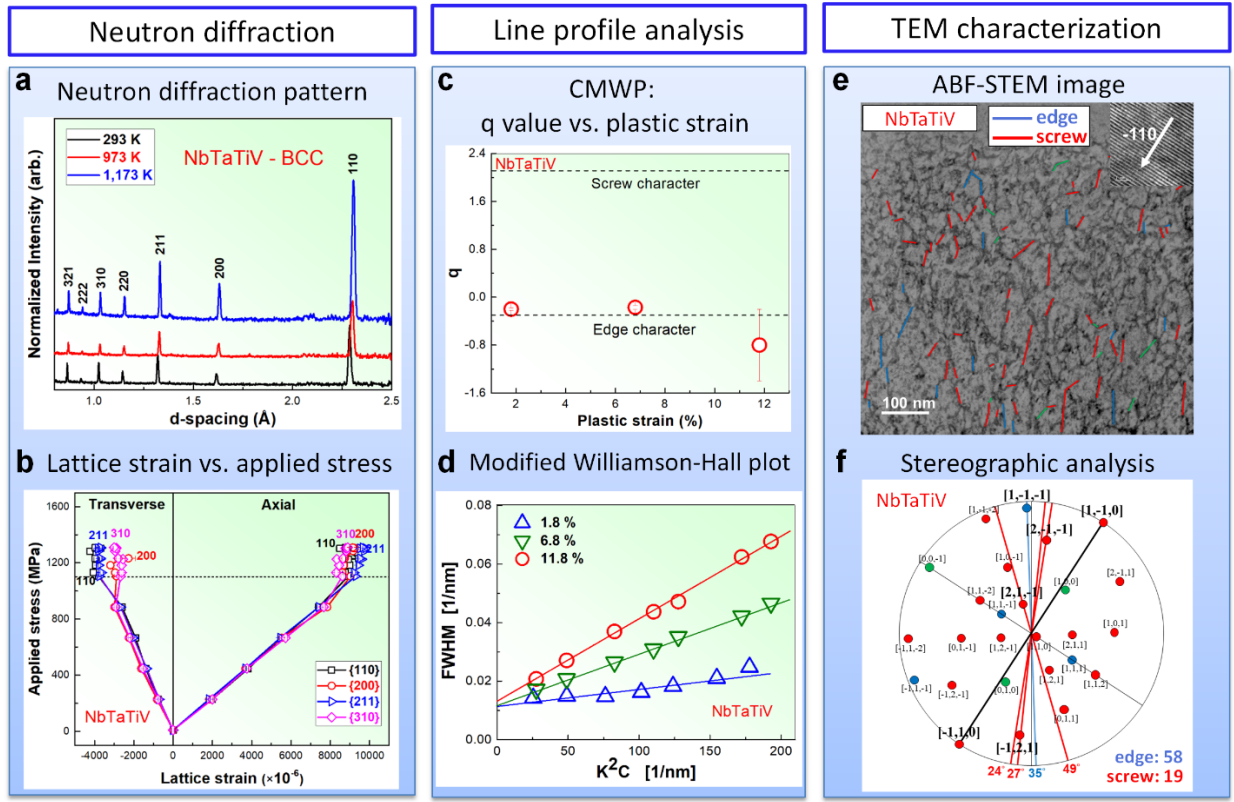

**Figure 2. Neutron diffraction and TEM experiments showing dominance of edge dislocations in NbTaTiV. (a)** Neutron-diffraction patterns showing interplanar spacings with peaks indexed for the BCC structure of NbTaTiV, at temperatures 293 K, 973 K, and 1,173 K. **(b)** Axial and transverse lattice strains versus applied load, shown as applied stress versus lattice strain so that the slopes correspond to the planar Young's moduli ($E_{hkl}$) and Young's moduli/Poisson's ratio ($E_{hkl}/\nu_{hkl}$), as indicated for the {110}, {200}, {211}, and {310} planes, respectively, at T = 293 K. The onset of plastic yielding (yield stress) is presented as the dashed line. **(c)** Evolution of q parameters as a function of plastic strain, which are obtained from Convolutional Multiple Whole Profile (CMWP) fitting. Dashed lines indicate values of q parameter for edge and screw character dislocations, considering 15 % error margin for the elastic constants. **(d)** Modified-Williamson-Hall plot, FWHM versus $K^2C$ at plastic strains of 1.8 %, 6.8 %, and 11.8 %, respectively, at T = 293 K. The plots were obtained from the physical profiles calculated by the CMWP procedure, considering a free of the instrumental effects on FWHM data. The pattern of the undeformed specimen was applied to CMWP procedure as an instrumental pattern. The much better agreement of the data with the edge analysis demonstrates the dominance of the edge dislocations. **(e)** Annular-Bright-field (ABF)-STEM image of NbTaTiV at 11.8 % plastic strains with two beam condition near $Z = [1\bar{1}3]$ and $\vec{g} = (110)$. All straight dislocation lines longer than 5 nm are highlighted by blue and red lines, corresponding to their identification as edge and screw dislocations, respectively. **(f)** Stereographic projection related to the [110] orientation, where [110] has been aligned with images in (e). All possible dislocations are indicated, and those corresponding to the images in (e) are highlighted in bold. The degrees indicate the angle with respect to the [110] direction. Blue lines/blue symbols indicate those dislocations identified as edge and red lines/symbols indicate those dislocations identified as screw. As a result, ~ 75% of the dislocations are identified as edge.

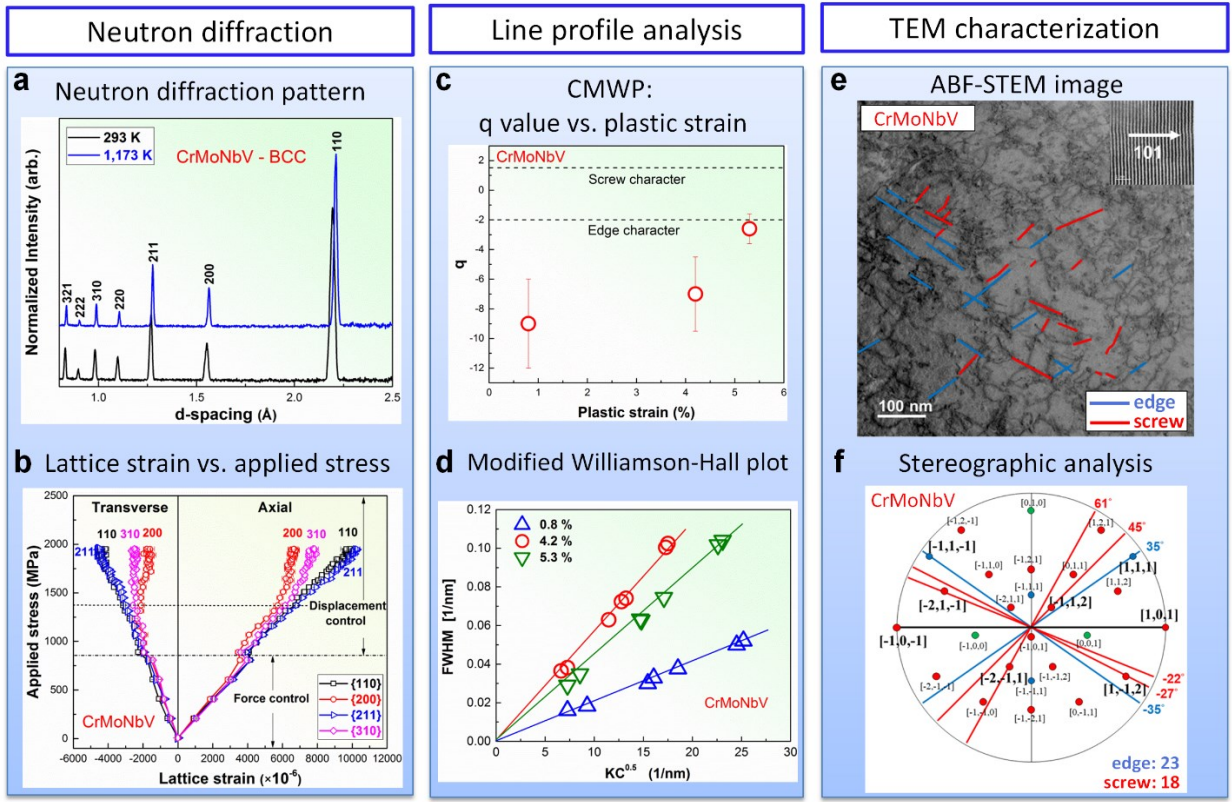

**Figure 3. Neutron diffraction and TEM experiments showing dominance of edge dislocations in CrMoNbV. (a)** Neutron-diffraction patterns showing interplanar spacings with peaks indexed for the BCC structure of CrMoNbV, at temperatures 293 K and 1,173 K. **(b)** Axial and transverse lattice strains versus applied load, shown as applied stress versus lattice strain so that the slopes correspond to the planar Young's moduli ($E_{hkl}$) and Young's moduli/Poisson's ratio ($E_{hkl}/\nu_{hkl}$), as indicated for the {110}, {200}, {211}, and {310} planes, respectively, at T = 293 K. The onset of plastic yielding (yield stress) is presented as the dashed line. **(c)** Evolution of q parameters as a function of plastic strain, which are obtained from Convolutional Multiple Whole Profile (CMWP) fitting. Dashed lines indicate values of q parameter for edge and screw character dislocations, considering 15 % error margin for the elastic constants. **(d)** Modified-Williamson-Hall plot, FWHM versus $KC^{0.5}$ at plastic strains of 0.8 %, 4.2 %, and 5.3 %, respectively, at T = 293 K. The plots were obtained from the physical profiles calculated by the CMWP procedure, considering a free of the instrumental effects on FWHM data. The pattern of the undeformed specimen was applied to CMWP procedure as an instrumental pattern. The much better agreement of the data with the edge analysis demonstrates the dominance of the edge dislocations. **(e)** Annular-Bright-field (ABF)-STEM image of CrMoNbV at 4.2 % plastic strains with two beam condition near $Z = [1\bar{1}3]$ and $\vec{g} = (110)$. All straight dislocation lines longer than 5 nm are highlighted by blue and red lines, corresponding to their identification as edge and screw dislocations, respectively. **(f)** Stereographic projection related to the [110] orientation, where [110] has been aligned with images in (e). All possible dislocations are indicated, and those corresponding to the images in (e) are highlighted in bold. The degrees indicate the angle with respect to the [110] direction. Blue lines/blue symbols indicate those dislocations identified as edge and red lines/symbols indicate those dislocations identified as screw. As a result, ~ 55% of the dislocations are identified as edge.

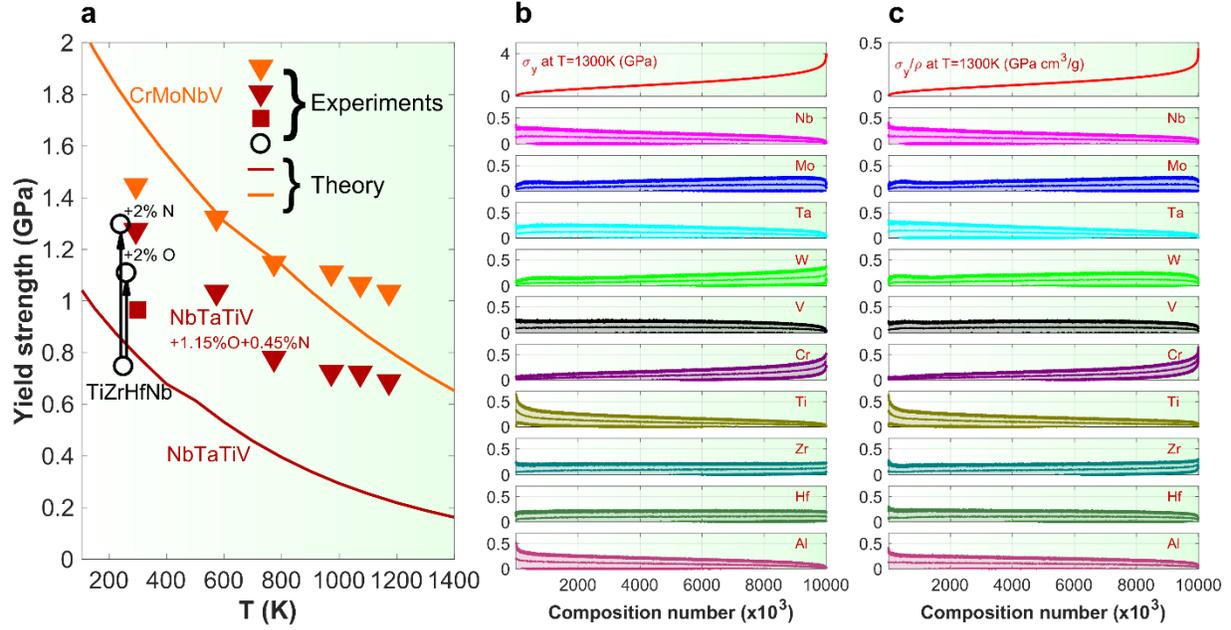

**Figure 4. Theory predictions of yield strength in BCC HEAs. (a)** Yield strength vs. temperature, experiments and theory for NbTaTiV and CrMoNbV. *NbTaTiV:* Experiments on NbTaTiV alloy with 1.15% O and 0.45% N: red triangles at a strain rate $10^{-3}$ s$^{-1}$; theory for *interstitial-free* alloy: red line. Red square reported in Ref. *(9)* at 5 x $10^{-4}$ s$^{-1}$ with no O or N impurity content reported. Black circles: experiments on TiHfZrNb at T = 300 K with and without 2% O and N that show strength increase up to ~ 500 MPa *(19)*, comparable to the difference here between model predictions for interstitial-free NbTaTiV and the experiments with 1.6% impurities. *CrMoNbV:* Experiments: orange triangles; theory: orange line; no impurities are detected in this alloy. The strength of the new CrMoNbV at 1,173 K exceeds those of all previous reported alloys. **(b,c)** Theory predictions for T = 1,300 K strength *vs* composition (b) and T = 1,300 K strength/weight ratio *vs* composition (c). The compositions indicate the average +/- standard deviation for 1,000 compositions per bin. The screening includes >10,000,000 compositions in the Nb-Mo-Ta-W-V-Cr-Ti-Zr-Hf-Al compositional space.

**Edge Dislocations Can Control Yield Strength**

**in Refractory Body-Centered-Cubic High Entropy Alloys**

Supplementary Materials

Francesco Maresca[1,2*,§], Chanho Lee[3,§], Rui Feng[3], Yi Chou[4], T. Ungar[5], Michael Widom[6],

Ke An[7], John D. Poplawsky[8], Yi-Chia Chou[4], Peter K. Liaw[3],

and W. A. Curtin[1]

1. Laboratory for Multiscale Mechanics Modeling, École Polytechnique Fédérale de Lausanne, CH-1015 Lausanne, Switzerland
2. Faculty of Science and Engineering, University of Groningen, Groningen, 9474AG, Netherlands
3. Department of Materials Science and Engineering, The University of Tennessee, Knoxville, TN 37996-2100, USA
4. Department of Electrophysics, National Chiao Tung University, Hsinchu, 30010, Taiwan
5. Department of Materials Physics, Eötvös University, Budapest, P.O. Box 32, H-1518, Hungary
6. Department of Physics, Carnegie Mellon University, Pittsburgh, PA 15213, USA
7. Neutron Scattering Division, Oak Ridge National Laboratory, Oak Ridge, TN, 37831, USA
8. Center for Nano-phase Materials Sciences, Oak Ridge National Laboratory, Oak Ridge, TN, 37831, USA

**Materials and Methods**

The NbTaTiV alloys were manufactured by arc-melting under the argon atmosphere on a water-cooled Cu hearth from Nb, Ta, Ti, and V elements of 99.99 weight percent (wt%) purity. The nominal composition of the present alloy is $Nb_{25}Ta_{25}Ti_{25}V_{25}$ in atomic percent (at%). To ensure a full synthesis of composed elements, the ingot was melted over 10 times. From the master alloys, the specimens were fabricated, followed by direct casting into cylindrical rods with a 4-mm diameter and 50-mm length, using a drop-casting technique. Similarly, the CrMoNbV alloy was fabricated by arc-melting the constituent elements of Cr, Mo, Nb, and V (purity 99.9 wt%), followed by drop casting into a water-cooled copper hearth. The NbTaTiV alloys were sealed with the triple-pumped argon in quartz tubes and homogenized at 1,473 K for 3 days, followed by water cooling. The CrMoNbV samples were tested and characterized in the as-cast state, due to the very high homogenous temperature (> 1,673 K). The microstructure was examined by scanning-electron microscopy (SEM), using a Zeiss Auriga 40 equipped with back-scattered electrons (BSE). The chemical composition of the alloy was studied by the atom probe tomography (APT) analysis after the homogenization treatment.

The mechanical tests were performed under uniaxial compression at elevated temperatures, employing a computer-controlled MTS servo-hydraulic-testing machine. Tests on NbTaTiV were conducted at an initial strain rate of $1 \times 10^{-3}$ s$^{-1}$ on specimens of 4 mm diameter and 8 mm length. Tests on CrMoNbV were conducted at a strain rate of $2 \times 10^{-4}$ s$^{-1}$ on specimens of 3 mm diameter and 6 mm length. Test samples were heated to and held at the desired temperatures for at least 30 min. until the temperature was stabilized within ± 10 K.

The neutron diffraction (ND) instrument utilizes time-of-flight (TOF) measurements, which allows covering a wide range of d spacings without the rotation of detectors or samples.

Two detectors at +/-90 degrees are used to collect diffracted beams from the polycrystalline grains with lattice planes parallel to the axial and transverse directions, respectively. The compression experiments on the homogenization-treated NbTaTiV alloy with a diameter of 4 mm and length of 8 mm and the as-cast CrMoNbV alloy with a diameter of 4.5 mm and length of 9 mm were conducted to investigate the elastic and plastic deformation behaviors at room and elevated temperatures, using a Materials Testing System (MTS) load frame. The samples are illuminated by the incident neutron beam of a $3 \times 3$ mm$^2$ slit size and 2-mm receiving collimators. During the measurement of the diffraction patterns for the elastic-deformation period, a stepwise-force control sequence was utilized. At each stress level, the measurement times of the ND data were 20 minutes and 12 minutes for NbTaTiV and CrMoNbV, respectively. For NbTaTiV, when the stress level reached 1,100 MPa for 293 K, 750 MPa for 973 K, and 620 MPa for 1,173 K (close to the macroscopic yield strength), a stepwise displacement control with an incremental step of 0.2 mm was employed. Similarly, for the CrMoNbV alloy, a stepwise displacement control with a displacement rate of $1.5 \times 10^{-4}$ mm/s was also used when the stress exceeds 800 MPa. The collected data were analyzed by single-peak fitting, using the VULCAN Data Reduction and Interactive Visualization software (VDRIVE) program *(26)*.

The TEM and STEM images of NbTaTiV and CrMoNbV were taken and analyzed, using JEOL ARM200F TEM/STEM with spherical aberration correctors. Stereographic projections of dislocations were used to identify the dislocation types and orientations.

**Supplementary Materials 1. Microstructure and chemical composition of the homogenized NbTaTiV HEA**

The Supp. Fig. 1 shows the scanning-electron microscopy (SEM) back-scattered electrons (BSE) images of the homogenized NbTaTiV HEA. The simple solid-solution microstructure was clearly observed with no formation of second phases or elemental segregation within grains and grain boundaries. The chemical composition of the homogenization-treated sample was identified by the APT analysis, which is close to the nominal composition ($Nb_{24.778}Ta_{23.471}Ti_{24.893}V_{24.929}C_{0.072}Ga_{0.278}Al_{0.002}N_{0.424}O_{1.153}$ on the average of the atom fractions). The CrMoNbV alloy shows a typical dendritic and interdendritic microstructure with a body-centered-cubic (BCC) structure, resulting from the different melting points of constituent elements. The detailed microstructure information of the CrMoNbV alloy can be found in a previous work *(24)*.

**Supplementary Materials 2. In-situ neutron diffraction and mechanical testing**

The lattice strains ($\varepsilon_{hkl}$) were calculated from the variation of the diffraction peak position, during loading by the following equation:

$$\varepsilon_{hkl} = \frac{d_{hkl} - d_{hkl}^0}{d_{hkl}^0} \quad \text{(S.1)}$$

where $d_{hkl}$ is the $hkl$ lattice spacing as a function of the applied stress, and $d_{hkl}^0$ is the reference $hkl$ lattice spacing under the unloaded state. The lattice-strain evolution can provide the $\{hkl\}$ plane-specific lattice strain during deformation. The evolution of lattice strains in the elastic region exhibits that the change of lattice strains is not dependent on the grain orientation, i.e., all oriented-grains [$\{110\}, \{200\}, \{211\}$, and $\{310\}$] present almost the similar elastic lattice strain with an identical level of the applied stress and maintain the linear response, indicating the elastic isotropy.

Supp. Fig. 2 exhibits the compressive engineering stress-strain curves for the homogenized NbTaTiV HEA at elevated temperatures. This alloy indicates the high yield strength ($\sigma_y$) of 1,239 MPa without the occurrence of fracture until the compressive strain of 30% at 293 K. As the temperature is increased to 1,173 K, the yield strength gradually reduces to 688 MPa with the maintenance of the excellent compressive plasticity. However, work softening is clearly observed when the sample is plastically deformed at 1,173 K. Supp. Fig. 3 displays the compressive engineering stress-strain curves for the as-cast CrMoNbV HEA at 293 K, 973 K, and 1,173 K. One can notice that this alloy exhibits a very high yield strength at room temperature (1,447 MPa). The yield strength of this alloy retains a high value (> 1,000 MPa) up to 1,173 K. The detailed values of yield strengths for the homogenized NbTaTiV and as-cast CrMoNbV alloys at 293 K, 973 K, and 1,173 K are noted in the Supp. Figs. 2 and 3.

**Supplementary Materials 3. TEM Analysis of Burgers vector and character in NbTaTiV**

The TEM contrast of a dislocation is related to **g·b** and when |**g·b**|/gb >1/3 the contrast is visible in the TEM. Supp. Table 1 summarizes relevant possible combinations of Burgers vectors and their contrast with respect to the **g** vectors were $[2\bar{2}\bar{2}]$, $[1\bar{1}\bar{2}]$ and $[1\bar{1}2]$. selected for our investigation. For all <100> Burgers vectors, |**g·b**|/gb>1/3 and such dislocations would show contrast in all the images. For some <110> and <111> Burgers vectors, |**g·b**|/gb<1/3 for some **g**, highlighted as red in the Table. Furthermore, both Burgers vectors $[1\bar{1}1]$ and $[101]$ have the same contrast conditions (highlighted by gray shading in the Table) and cannot be distinguished. Furthermore, both Burgers vectors $[1\bar{1}\bar{1}]$ and $[10\bar{1}]$ have the same contrast conditions (highlighted by blue in the Supp. Table 1) and so also cannot be distinguished. Only the $[111]$ Burgers vector showed dislocation contrast for **g** of $[1\bar{1}\bar{2}]$ and $[1\bar{1}2]$, and invisibility for $[2\bar{2}\bar{2}]$. The $[110]$ Burgers vector is the only case that shows no dislocation contrast for all **g** vectors. Based on these various conditions, we can interpret our STEM data showing the dislocation contrast for the various imaging conditions (**g** vectors). Not all dislocations can be uniquely determined but our subsequent stereographic projection based on <111> Burgers vectors provides complementary data that the dislocations have the <111> Burgers vectors.

Supp. Fig. 4 shows the scanning transmission electron microscopy (STEM) annular dark field (ADF) images viewed along the [110] direction for different conditions. Regions A, B, and C are indicated only for reference in the discussion. In Supp. Fig. 4a, all diffraction spots are included and hence all dislocations are visible. Supp. Fig. 4b showed the STEM ADF image for the $[2\bar{2}\bar{2}]$ g vector, and many of the dislocations in Region A are now invisible. In region B, nearly horizontal dislocation lines appear. In region C, the contrast is blurred but the top-half area is grayer and shows less dislocation contrast. Supp. Fig. 4c has the $[1\bar{1}\bar{2}]$ g vector, and region A shows two different contrasts, with the top area and bottom-left corner showing lighter contrast, and the center area showing darker contrast. The bottom-half area in region B and the top-half area in region C also show contrast. Supp. Fig. 4d has the $[1\bar{1}2]$ g vector, and region A shows lighter contrast in the top half and the bottom-left corner while region C shows lighter contrast in the top half with obvious vertical dislocation lines.

According to Supp. Table 1, the contrast conditions observed for the top and bottom-left-corner in region A and for the top half in region C are only in agreement with [111] Burgers vector (contrast for $[1\bar{1}2]$ and $[1\bar{1}2]$, no contrast for $[2\bar{2}\bar{2}]$). Region B provides information on the dislocation character as discussed below. The center area in region A matches with either $[1\bar{1}1]$ or $[10\bar{1}]$ Burgers vectors (contrast for $[1\bar{1}2]$, no contrast for $[2\bar{2}\bar{2}]$ or $[1\bar{1}2]$). Therefore, only the <111> type Burgers vectors are consistent with the contrast in all regions for all g vectors. While $[10\bar{1}]$ is theoretically possible in region A, it is absent in other regions and would be an unusual Burgers vector for slip in a BCC crystal. We conclude that the Burgers vectors in the NbTaTiV alloy are of the <111> type that is the well-established Burgers vector in elemental and dilute BCC alloys.

The TEM or STEM images can provide some further indication regarding dislocation character, as follows. Recall that the line directions for edge and screw dislocation are perpendicular and parallel to **b**, respectively. Therefore, if the g vector is parallel to the dislocation line, screw dislocations will have the largest contrast and edge dislocations will not be visible. If the projection of the dislocation line onto the [110] plane (axis of viewing) is aligned with **g**, then the contrast can determine the character. In addition, if the g vector is perpendicular to the dislocation line, screw dislocation will not show any dislocation contrast but edge dislocation may show dislocation contrast when $|\mathbf{g}\cdot\mathbf{b}|/gb<1/3$.

For example, a $[1\bar{1}\bar{1}]$ screw dislocation has a $[1\bar{1}\bar{1}]$ line direction while a $[111]$ edge dislocation has a $[2\bar{1}\bar{1}]$ line direction. The projections of these two dislocations onto the (110) plane leads to projected line directions that differ by only 10.02 degrees while the angle between the Burgers vectors is 70.52 degrees. Looking closely at Region A in Supp. Fig. 4a, we can observe a predominance of near-vertical dislocation lines. If these near-vertical dislocation line are a screw type, the Burgers vector should be $[1\bar{1}\bar{1}]$ and on the (110) plane. For g=$[2\bar{2}\bar{2}]$, which is near vertical, such a $[1\bar{1}\bar{1}]$ screw dislocation should be clearly visible. If the line is invisible, then the dislocation must be of edge character, which is the situation seen in region A for Supp. Fig. 4b.

As another example, near-$[1\bar{1}2]$ dislocation lines in region B of Supp. Fig. 4b with **g**=$[2\bar{2}\bar{2}]$ are observed. A screw dislocation should show no contrast but here the dislocation lines are clearly visible. Similarly, near $[1\bar{1}2]$ and $[2\bar{1}\bar{1}]$ (the projected vector is $[3\bar{3}\bar{2}]$ on (110) plane), dislocation lines are observed in Supp. Fig. 4d with g=$[1\bar{1}2]$ and match with edge character.

The **g·b** contrast cannot reveal the character of all the observed dislocation. The stereographic projection method is thus used for better determination of the overall distribution of dislocation character, as shown in the main text.

| b | | $g_{[2\bar{2}\bar{2}]}\cdot b$ | $g_{[1\bar{1}\bar{2}]}\cdot b$ | $g_{[1\bar{1}2]}\cdot b$ |
|---|---|---|---|---|
| <111> | [111] | -0.33 | -0.47 | 0.47 |
|  | [1$\bar{1}$1] | 0.33 | 0.00 | 0.94 |
|  | [1$\bar{1}\bar{1}$] | 1.00 | 0.94 | 0.00 |
| <110> | [110] | 0.00 | 0.00 | 0.00 |
|  | [101] | 0.00 | -0.29 | 0.87 |
|  | [10$\bar{1}$] | 0.82 | 0.87 | -0.29 |
|  | [1$\bar{1}$0] | 0.82 | 0.58 | 0.58 |
| <100> | [100] | 0.58 | 0.41 | 0.41 |
|  | [010] | -0.58 | -0.41 | -0.41 |
|  | [001] | -0.58 | -0.82 | 0.82 |

**Supplementary Table 1.** Normalized g·b values for possible Burgers vectors under the imaging conditions used in Supp. Fig. 4. Absolute values |**g·b**|/gb<1/3 are invisible and are highlighted by red. Pairs of Burgers vectors with the |**g·b**|/gb shaded in either gray or blue are not distinguishable under these conditions.

**Supplementary Materials 4. Analysis of elastic response**

Supp. Fig. 5 shows the reciprocal diffraction elastic constants ($1/E_{hkl}$ and $v_{hkl}/E_{hkl}$), calculated by the Kröner model, plotted as a function of the elastic-anisotropy factor, $A_{hkl} = \left\{\frac{h^2k^2+k^2l^2+l^2h^2}{(h^2+k^2+l^2)^2}\right\}$ at both room and elevated temperatures *(27-29)*. It is found that the theoretical Kröner model fits the experimental data very well at all temperatures. Based on this good agreement with the in-situ neutron experimental data, the single-crystal elastic constants ($C_{ij}$), macroscopic bulk modulus ($K_M$), and shear modulus ($G_M$) were calculated, using Kröner's self-consistent model *(28,29)*:

$$G_K^3 + \alpha G_K^2 + \beta G_K + \gamma = 0 \qquad (2)$$

where $G_K$ is the diffraction shear modulus, and $\alpha, \beta$, and $\gamma$ are constants given by [S.5]:

$$\alpha = \frac{3\{3K_M + 4[\mu + 3(\eta-\mu)A_{hkl}]\}}{8} - \frac{(2\eta + 3\mu)}{5} \qquad (3)$$

$$\beta = \frac{3K_M[\mu + 3(\eta-\mu)A_{hkl}]}{4} - \frac{3(6K_M\eta + 9K_M\mu + 20\eta\mu)}{40} \qquad (4)$$

$$\gamma = -\frac{3K_M\eta\mu}{4} \qquad (5)$$

where $K_M$ is the bulk modulus, which is given by $(C_{11} + C_{12})/3$, $\eta$ is $(C_{11} - C_{12})/2$, and $\mu$ is equal to $C_{44}$. The value of the isotropic macroscopic shear modulus, $G_M$, becomes the diffraction shear modulus, $G_K$, in Eq. (2), by substituting 0.2 for $A_{hkl}$, if averaging over all orientations. Then, the isotropic macroscopic Young's modulus, $E_M$, can be calculated by the equation:

$$E_M = \frac{9G_M K_M}{G_M + 3K_M} \qquad (6)$$

The calculated single-crystal elastic constants ($C_{ij}$), macroscopic Young's ($E_M$), shear ($G_M$), bulk ($K_M$) moduli, and Poisson's ratio ($v$) at room and elevated temperatures are listed in the Supp. Fig. 5.

**Supplementary Materials 5. Finding high-temperature strengths in the whole Cr-Mo-Nb-Ta-V-W-Ti-Zr-Hf-Al composition space.**

Figures 4b,c show the predictions of the reduced theory (see the main text) as a function of the composition for > 10,000,000 compositions in the whole Cr-Mo-Nb-Ta-V-W-Ti-Zr-Hf-Al space. The whole compositional space is explored by varying the alloy content of each element in 5 at% steps. For the clarity of visualization, the compositions are binned in groups of 1,000 and the average +/- standard deviation is shown, per each element. The screening is performed by using as input the single-element atomic volumes, elastic constants, and densities, all listed in Supp. Table 2. The alloy values are then computed by using the rule of mixtures of the elemental values (see the main text).

For the case of Al, the atomic volume is 14,075 $Å^3$, based on the work by Chen et al. *(30)*. The atomic volumes of all other elements are the same as reported in *(31)*. For the Ti and Zr, the values are obtained by extrapolating high-temperature (high-T) measurements to room temperature (RT), while Hf is obtained by using the Vegard's law on Hf-HEAs. The atomic volumes adopted for Ti, Zr, and Hf are similar to those estimated in Ref. *(21)*, which were instead obtained by extrapolating the elemental values from binary alloys in the literature.

The cubic elasticity constants of the BCC Al are assumed to be equal to the FCC values. The Ti, Zr, and Hf values are taken from high-T phonon measurements (at 1,293 K, 1,188 K, and 2,073 K, respectively), see *(32-34)*.

|     | Volume (Å³) | $C_{11}$ (GPa) | $C_{12}$ (GPa) | $C_{44}$ (GPa) |
| --- | --- | --- | --- | --- |
| **Al** | 14.075 | 105.6 | 63.9 | 28.53 |
| **Cr** | 12.321 | 339.8 | 58.6 | 99 |
| **Hf** | 22.528 | 131 | 103 | 45 |
| **Mo** | 15.524 | 450.02 | 172.92 | 125.03 |
| **Nb** | 17.952 | 252.7 | 133.2 | 30.97 |
| **Ta** | 17.985 | 266.32 | 158.16 | 87.36 |
| **Ti** | 17.387 | 134 | 110 | 36 |
| **V** | 14.020 | 232.4 | 119.36 | 45.95 |
| **W** | 15.807 | 532.55 | 204.95 | 163.13 |
| **Zr** | 23.02 | 104 | 93 | 38 |

**Supplementary Table 2.** Single-element properties used for theoretical predictions.

**Supplementary Materials 6. Energetic competition within the Cr-Mo-W-Zr alloy system**

In order to achieve a BCC solid solution we require the free energy to be lower than competing phases. Two factors enter consideration, enthalpy and entropy. We will compute enthalpies within the Cr-Mo-W-Zr quaternary alloy system utilizing density functional theory (DFT). In addition to the pure elements and random solid solutions, the competing phases are ordered BCC of Pearson type cP2 (B2) and Laves phases of Pearson types hP12, hP24, and cF24 (Strukturbericht C14, C36, and C15), and $Mo_3Zr.cP8$ and $W_5Zr_3.tI32$.

Our DFT calculations employ the plane-wave code VASP *(35)* in the PBE generalized gradient approximation *(36)* with an energy cutoff of 300 eV and k-point densities sufficient to converge energies to within 1 meV/atom. We fully relax atomic positions and lattice parameters, and consider antiferromagnetic spin polarization in the case of elemental Cr. Enthalpies are calculated relative to pure elements. Stable phases lie on the convex hull of enthalpies. Instability energies $\Delta E$ are defined as the height above the convex hull *(37)*.

BCC solid solutions are represented as randomly occupied supercells containing $N_1 - N_4$ atoms of the four different species. A total of

$$\Omega(N_1, N_2, N_3, N_4) = \frac{(N_1 + N_2 + N_3 + N_4)!}{N_1! \, N_2! \, N_3! \, N_4!}$$

such configurations exist from which we take $N_{samples} = 20$ representative configurations in order to obtain a distribution of instability energies $\{\Delta E_k\}$. Our energies $\Delta E_k$ range from 232 meV/atom up to 314, with an average value of 266 and standard deviation of 15. The energies define a partition function *(38)*

$$Z = \frac{\Omega(N_1, N_2, N_3, N_4)}{N_{samples}} \sum_{k=1}^{N_{samples}} e^{-E_k/k_B T}$$

and its associated free energy, $F = -k_B T ln(Z)$, in which varying degrees of short-range chemical order are appropriately weighted. The BCC solid solution gains stability relative to competing phases above the temperature $T_0$ at which $F$ vanishes. Vibrational and electronic entropy are neglected, as these are found to be small effects relative to the entropy of chemical substitution when comparing phases of similar structure (*i.e.* BCC) *(39)*.

For equiatomic CrMoWZr we predict phase separation at low temperature into a mixture of BCC-Cr, $Cr_2Zr.cF24$, $W_2Zr.cF24$, and BCC-Mo. The transition to single phase BCC occurs at $T_0$=2300K. The melting temperature is not precisely known; we estimate it as 2300K by averaging the melting temperatures of the six equiatomic binaries. Thus we predict the equiatomic BCC phase to be unstable at all temperatures below melting. Because Laves phase are major competitors to the HEA, we estimated the free energy of the cF24 Laves phase assuming concentration $Mo_2Zr_6$ on site 8a and $Cr_6Mo_4W_6$ on site 16d. This structure lies 82 meV/atom above the convex hull, suggesting that it could be stabilized by entropy of mixing on the sublattices above $T_0$=1300K.

Cr and Zr are especially prone to Laves phase formation, so we investigated the effect of moving off-stoichiometry, to $CrMo_2W_2Zr$ within a 24-atom supercell. Because the composition moves away from the $Cr_2Zr$ and $W_2Zr$ Laves phases, the distribution of $\Delta E_k$ values shifts downward by approximately 100 meV/atom, and we predict formation of single phase BCC at $T_0$ = 1600K. At the same time, the melting temperature should rise because the composition is enriched in elements Mo and W whose melting temperatures are high. Thus we obtain a thermally stable non-stoichiometric HEA over a wide temperature range. However, this analysis considers only a subset of potentially competing phases that omits binary and ternary solid solutions, so further investigation will be required to validate this prediction.

The situation is similar for CrHfMoW, which phase separates at low temperature into a mixture of BCC phases plus the Laves phase $HfW_2$. The transition to single phase BCC occurs at $T_0$=2400K, compared with our estimated melting temperature of 2100K. The high entropy Laves phase is stabilized above $T_0$ = 1350K. At composition $CrHfMo_2W_2$, the predicted $T_0$ drops to 1200K, while the melting temperature rises.

**Supplementary References**

26. K. An, VDRIVE-Data reduction and interactive visualization software for event mode neutron diffraction, ORNL Report No. ORNL-TM-2012-621 Oak Ridge National Laboratory, Oak Ridge, TN (2012).

27. E. Kröner, Berechnung der elastischen Konstanten des Vielkristalls aus den Konstanten des Einkristalls. *Z. Phys.* **151** 504-518 (1958).

28. R. De Wit, Diffraction elastic constants of a cubic polycrystal. *J. Appl. Crystallogr.* **30**, 510-511 (1997).

29. T. Gnäupel-Herold, P.C. Brand, H.J. Prask, Calculation of single-crystal elastic constants for cubic crystal symmetry from powder diffraction data. *J. Appl. Crystallogr.* **31**, 929-935 (1998).

30. H. Chen, A. Kauffmann, S. Laube, I.-C. Choi, R. Schwaiger, Y. Huang, K. Lichtenberg, F. Müller, B. Gorr, H.-J. Christ, M. Heilmaier, Contribution of lattice distortion to solid solution strengtheninng in a series of refractory high entropy alloys. *Metall. Mater. Trans. A* **49**, 772-781 (2018).

31. B. Yin, F. Maresca, W.A. Curtin, Vanadium is an optimal element for strengthening in both fcc and bcc high-entropy alloys. *Acta Mater.* **188**, 486–491 (2020).

32. W. Petry, A. Heiming, J. Trampenau, M. Alba, C. Herzig, H.R. Schober, G. Vogl, Phonon dispersion of the bcc phase of group-IV metals. I. bcc titanium. *Phys. Rev. B* **43**, 10933-10947 (1991).

33. A. Heiming, W. Petry, J. Trampenau, M. Alba, C. Herzig, H.R. Schober, G. Vogl, Phonon dispersion of the bcc phase of group-IV metals. II. bcc zirconium, a model case of dynamical precursor of martensitic transitions. *Phys. Rev. B* **43**, 10948-10962 (1991).

34. J. Trampenau, A. Heiming, W. Petry, M. Alba, C. Herzig, W. Miekeley, H.R. Schober, Phonon dispersion of the bcc phase of group-IV metals. III. bcc hafnium. *Phys. Rev. B* **43**, 10963-10969 (1991).

35. G. Kresse, J. Furthmuller, Efficient iterative schemes for ab initio total-energy calculations using a plane-wave basis set, *Phys. Rev. B* **54**, 11169 (1996).

36. J. P. Perdew, K. Burke, M. Ernzerhof, Generalized Gradient Approximation Made Simple, *Phys. Rev. Lett.* **77**, 3865 (1996).

37. M. Mihalkovic, M. Widom, Ab-initio calculations of cohesive energies of Fe-based glass-forming alloys, *Phys. Rev. B* **70**, 144107 (2004).

38. H. Zhang, S. Yao, M. Widom, Predicted phase diagram of B-C-N, *Phys. Rev. B* **93**, 144107 (2015).

39. M. Gao, P. Gao, J. Hawk, L. Ouyang, D. Alman, M. Widom, Computational modeling of High-Entropy Alloys: Structures, thermodynamics and elasticity, *J. Mater. Res.* **32**, 3627-3641 (2017).

# Supplementary Figures

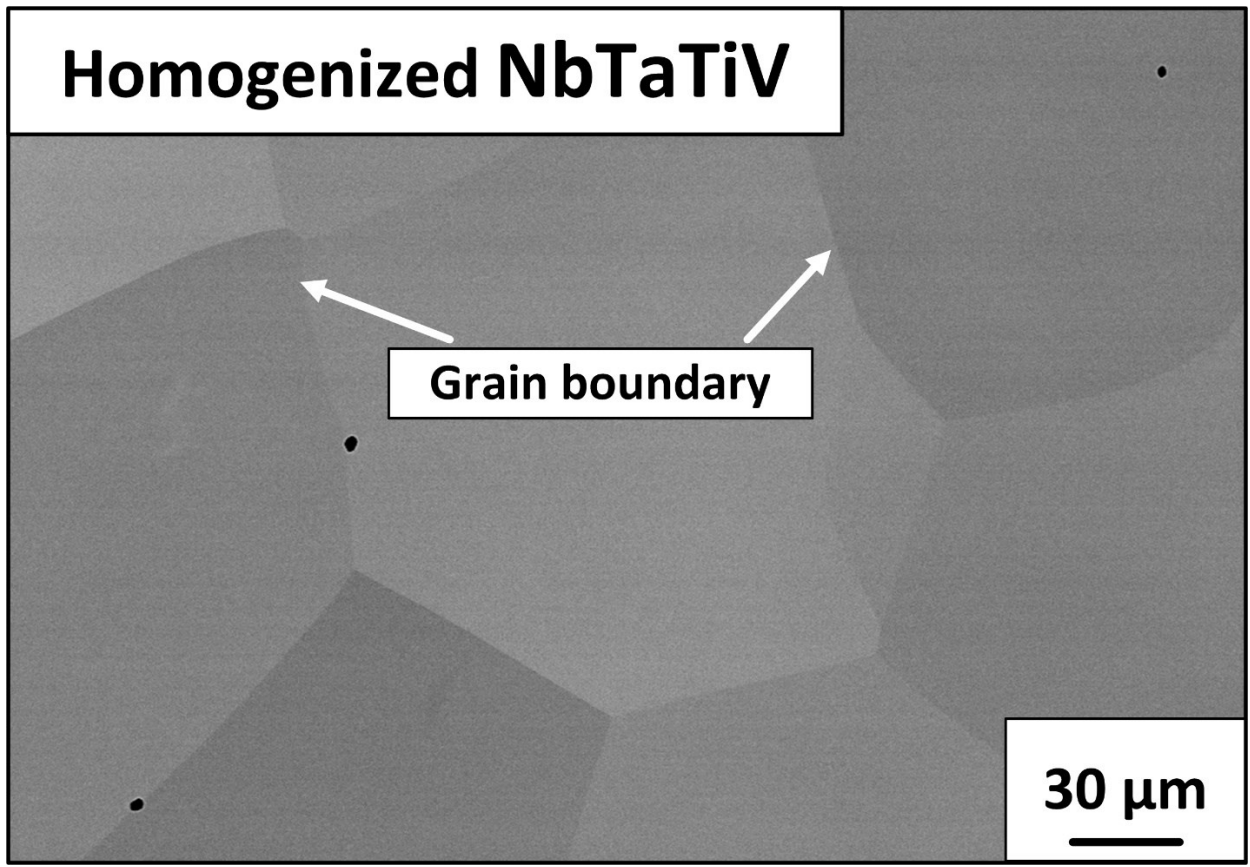

**Supplementary Figure 1.** SEM-BSE image of the homogenization-treated NbTaTiV refractory HEA.

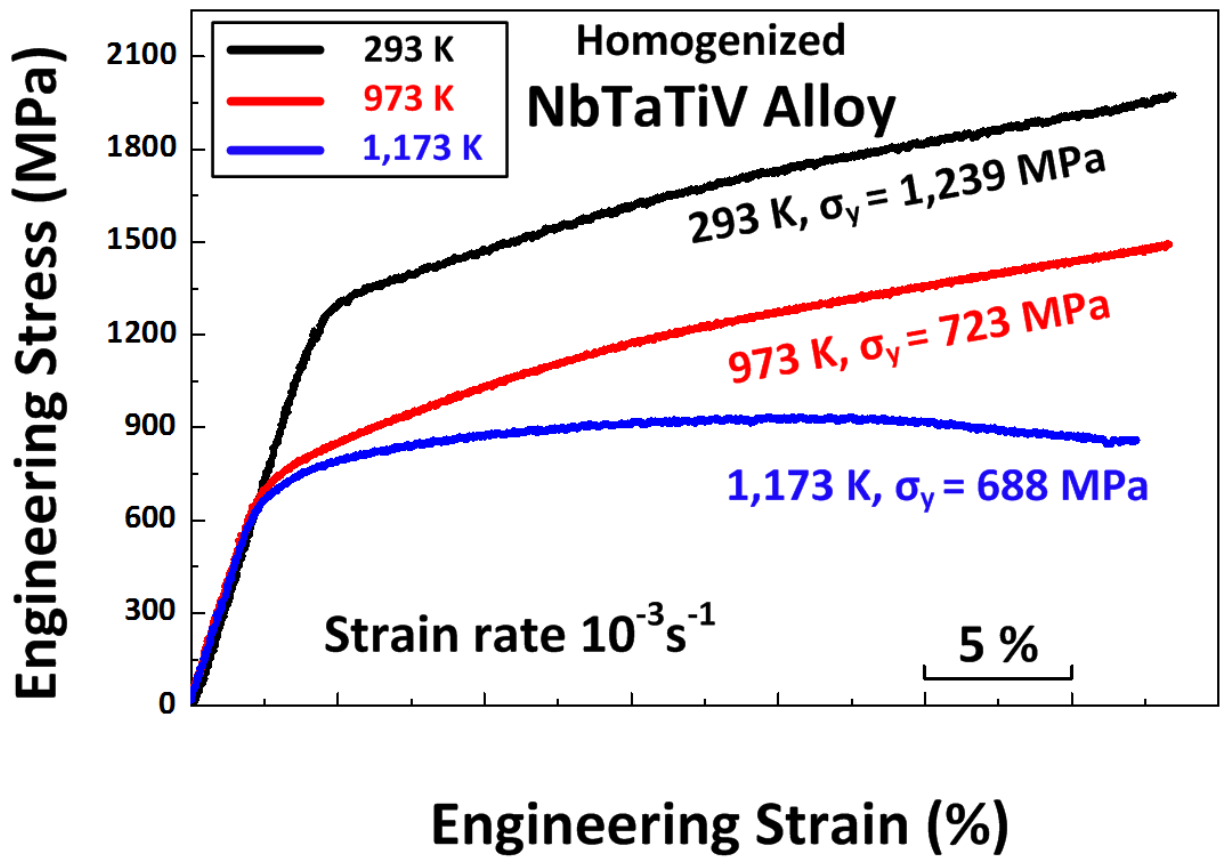

**Supplementary Figure 2.** Mechanical properties of the homogenized NbTaTiV HEA obtained at 293 K, 973 K, and 1,173 K, respectively.

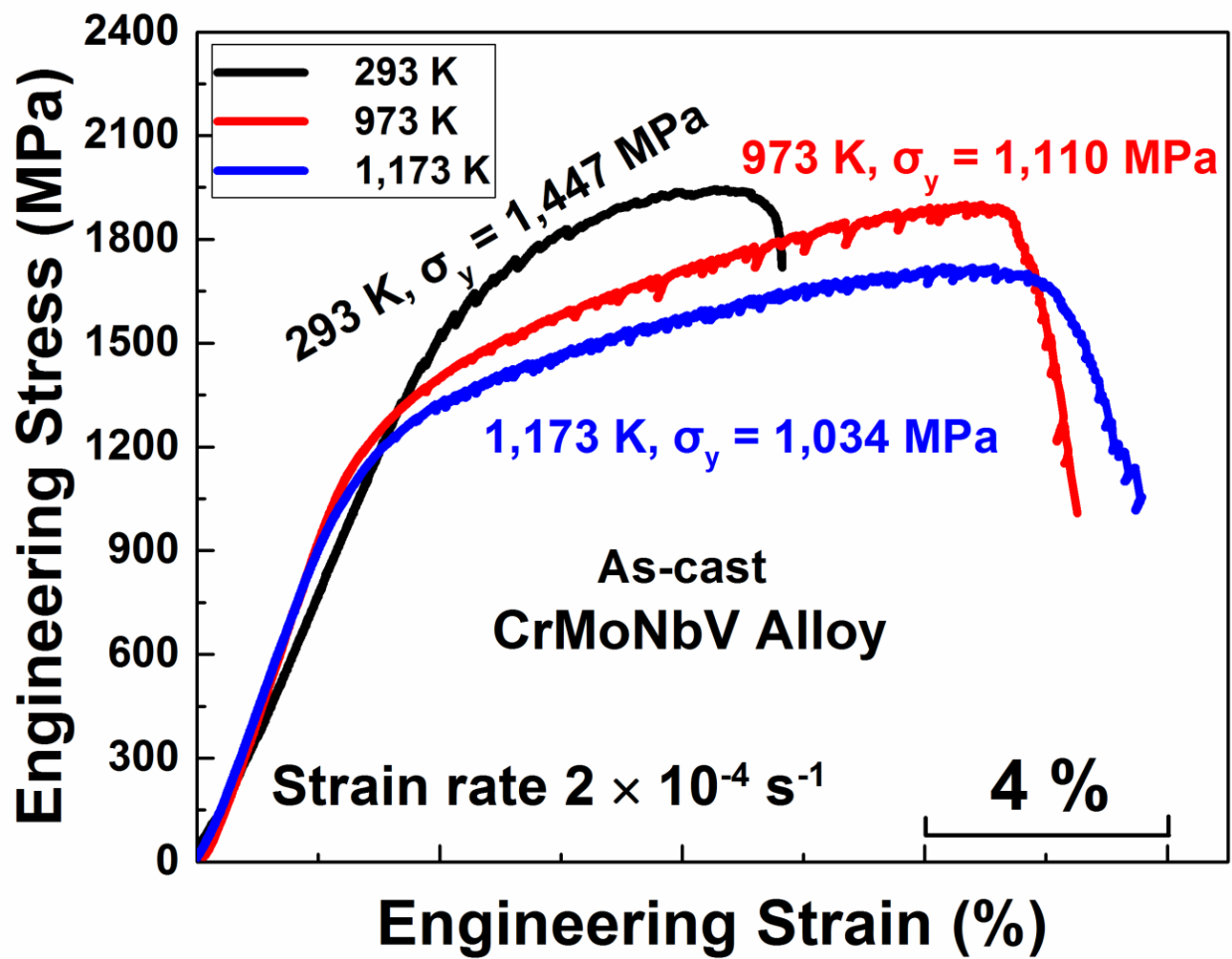

**Supplementary Figure 3.** Mechanical properties of the as-cast CrMoNbV HEA obtained at 293 K, 973 K and 1,173 K, respectively.

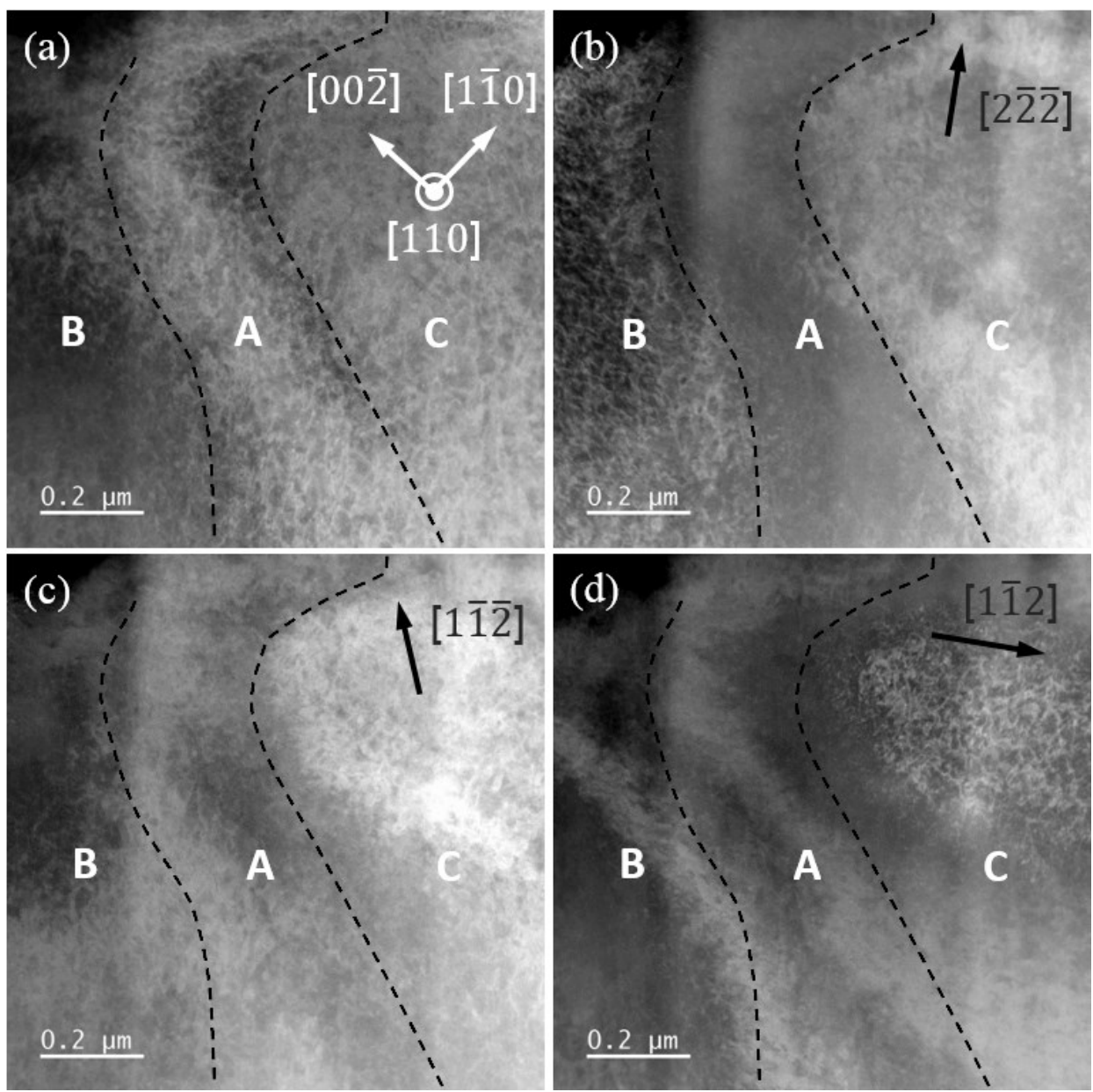

**Supplementary Figure 4.** The analysis on dislocation types in 15% deformed NbTaTiV. All the images were taken at the same region but different imaging conditions. The line marks separate the regions showed different contrast at each imaging condition. **(a)** ADF image viewing along [110] direction. The two lowest indexed direction were marked with arrows. The ADF images viewing along [110] and with **(b)** [2$\bar{2}\bar{2}$], **(c)** [1$\bar{1}\bar{2}$] and **(d)** [1$\bar{1}$2] g vectors. The g vectors were marked on the images with arrows.

**Plot of diffraction elastic constants of NbTaTiV HEA and fitting with Kroner model**

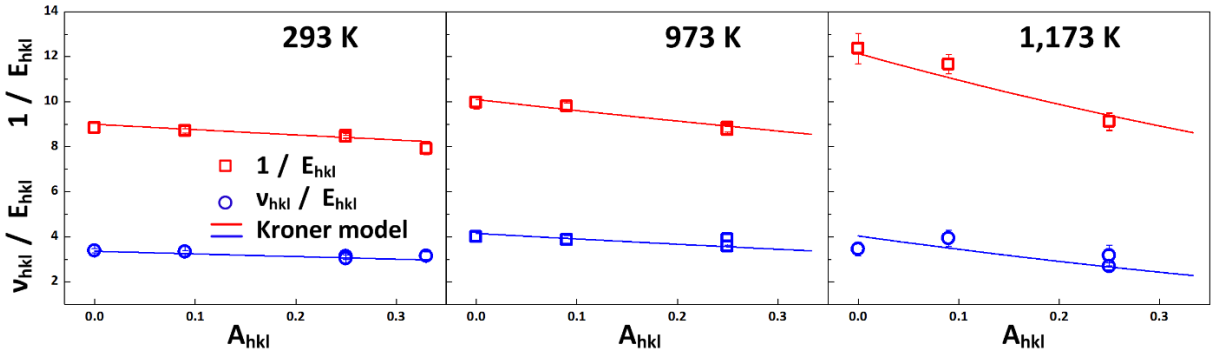

| Parameters / Conditions | $C_{11}$ (GPa) | $C_{12}$ (GPa) | $C_{44}$ (GPa) | $C'$ (GPa) | $E_M$ (GPa) | $G_M$ (GPa) | $K_M$ (GPa) | $v$ |
|---|---|---|---|---|---|---|---|---|
| 293 K | 196.79 | 121.44 | 46.74 | 37.68 | 117.3 | 42.9 | 146.6 | 0.368 |
| 943 K | 184.33 | 122.52 | 45.93 | 30.91 | 107.9 | 39.3 | 143.1 | 0.376 |
| 1,173 K | 134.87 | 76.78 | 46.65 | 29.05 | 102.3 | 38.7 | 96.1 | 0.325 |

**Supplementary Figure 5.** Reciprocal diffraction elastic constants, $1/E_{\{hkl\}}$ and $v_{hkl}/E_{\{hkl\}}$, as a function of $A_{hkl}$, and fitting with the Kröner model. The single-crystal macroscopic elastic constants, $C_{11}$, $C_{12}$, and $C_{44}$, the macroscopic Young's ($E_M$), shear ($G_M$), bulk ($K_M$) moduli, and Poisson's ratio ($v$) at 293 K, 973 K, and 1,173 K, respectively.